# Multi-objective Clustering Algorithm with Parallel Games

Dalila Kessira
*Laboratory of Medical Informatics (LIMED)*
*University of Bejaia*
Bejaia, Algeria
dalila.kessira@univ-bejaia.dz

Mohand-Tahar Kechadi
*Insight Center for Data Analytics*
*University College of Dublin*
Dublin, Ireland
tahar.kechadi@ucd.ie

*Abstract*— Data mining and knowledge discovery are two important growing research fields in the last two decades due to the abundance of data collected from various sources. The exponentially growing volumes of generated data urge the development of several mining techniques to feed the needs for automatically derived knowledge. Clustering analysis (finding similar groups of data) is a well-established and widely used approach in data mining and knowledge discovery. In this paper, we introduce a clustering technique that uses game theory models to tackle multi-objective application problems. The main idea is to exploit a specific type of simultaneous move games, called congestion games. Congestion games offer numerous advantages ranging from being succinctly represented to possessing Nash equilibrium that is reachable in a polynomial-time. The proposed algorithm has three main steps: 1) it starts by identifying the initial players (or the cluster-heads); 2) then, it establishes the initial clusters' composition by constructing the game and try to find the equilibrium of the game. The third step consists of merging close clusters to obtain the final clusters. The experimental results show that the proposed clustering approach obtains good results and it is very promising in terms of scalability and performance.

*Keywords*— Data mining, data analysis, clustering, game theory, simultaneous-move games, Nash equilibrium.

## I. Introduction

Data science is one of the most growing research fields over the last few years. It refers to an empirical approach that uses the available big amounts of data to provide answers to a wide variety of questions and problems that human beings are enable to treat without the intervention of computers [1]. Among the several methods and techniques used in data science, cluster analysis (clustering) is a well-established and widely used technique. It is the process of grouping the data into classes or clusters, in such a way that objects in the same cluster are very similar and objects in different clusters are different.

Game theory offers very attractive rigorous mathematical tools for modelling and resolving various strategic situations; thus, it is increasingly used to resolve a wide range of problems.

Game theory-based clustering algorithms are not numerous, and only a few of them do reach the Nash equilibrium in a polynomial time. This is due to the high complexity of finding solution concepts, such as Nash equilibrium [2]. There are many other types of clustering algorithms based on various methodologies and concepts (see [3][4], [5] , for more details).

In this paper, we propose a multi-objective clustering algorithm based on game theory using parallel games. It is called MOCA-SM (Multi-Objective Clustering Algorithm based on Simultaneous-Move games).

The rest of this paper is organized as follows: in section 2, we give an overview of Game Theory and its basic concepts. In section 3 we present the proposed clustering approach. In Section 4 we provide experimental results and discussion. In Section 5 we conclude and give some future directions and perspectives of this research work.

## II. Background

In this section, we first introduce the game theory, then, we present the subclass of games used in this work, which is called Singleton Congestion games with player-specific payoff functions.

### A. Parallel Games

In game theory, the term "game" means an abstract mathematical model of a multi-agent decision-making setting[6]. Formally, it is represented by a tuple (N, S, u) [7], where

- $N = \{1, ..., n\}$ is a set of players, indexed by i;
- $S = S_1 * ... * S_n$, and $S_i$ is a finite set of strategies available to player i. Each vector $s = (s_1, ..., s_n) \in S$ is called a strategy profile;
- $u = (u_1, ..., u_n)$, and $u_i: S \rightarrow R$ is a real-valued utility (or payoff) function for player i.

The normal form, also known as the strategic or matrix form is when we represent a game via an n-dimensional matrix. Simultaneous-move (parallel) games [8] are when all players take their decisions at the same time, consequently, they do not know the decisions made by the other players.

### B. Solution Concepts and Nash Equilibrium

Reasoning about multi-player games is based on solution concepts, which are principals that help us to identify interesting subsets of the outcomes of a game [7]. One of the most powerful solution concepts in game theory is Nash equilibrium.

To define Nash equilibrium [2], we need to define the notion of best response [7]. Formally, we denote $s_{-i} = (s_1, ..., s_{i-1}, s_{i+1}, ..., s_n)$, a strategy profile s without player i's strategy. Thus, we can write $s = (s_i, s_{-i})$. If the agents other than i (whom we denote −i) were to commit to playing $s_{-i}$, a utility-maximizing agent i would face the problem of determining their best response.

**Definition 1 (Best response) [7]:** Player i's best response to the strategy profile $s_{-i}$ is a mixed strategy $s^*_i \in S_i$ such that $u_i(s^*_i, s_{-i}) \geq u_i(s_i, s_{-i})$ for all strategies $s_i \in S_i$.

The best response is not necessarily unique. Thus, the notion of "best response" is not a solution concept; it does not identify an interesting set of outcomes in this general case [7].

**Definition 2 (Nash equilibrium) [2]:** A strategy profile $s = (s_1,..., s_n)$ is a Nash equilibrium if $s_i$ is a best response to $s_{-i}$ for all agents i.

Nash equilibrium is a simple but powerful principle for reasoning about behavior in general games [2]: even when there is no dominant strategy, we should expect players to use strategies that give the best responses to each other [9]. This is the most used solution concept in game theory, nevertheless, it is established that computing Nash equilibrium is hard for many games' subclasses [10]–[13].

*C. Singleton Congestion Games with Player-Specific Payoff Functions*

In 1996 Milchtaich introduced a subclass of congestion games which were presented by Rosenthal in [14] [15]. In this subclass, the payoff function associated with each resource is not universal but a player specific. And it is assumed that each player chooses only one primary resource and that the actual payoff received decreases (not necessarily strictly decreasing) with the number of other players who selected the same resource. Consequently, the cost of a resource *e* for player *i* does not depend only on the number of players that chose it, but it depends also on the player himself. Hence, every player *i* has their cost function $c_{ie}$ for every resource *e*, hence the introduction of the definition below.

**Definition 3 (SCGPSC:** Singleton congestion game with player-specific cost [14]) is a tuple (N, E, S, $(c_{ie})_{i \in N, e \in E}$), where
- N = {1, ..., n} is a set of players, indexed by i.
- E = {1, ..., m} is a set of resources, indexed by e.
- Strategy set S: $S_1 \times ... \times S_n$, where $S_i$ is a set of strategies available to player i and strategy $s_i \in S_i$ is a singleton; a set with exactly one element.
- Cost function $c_{ie}(n_e) \in \mathbb{R}$ for player i and resource e, where $c_{ie}(n_e) \neq c_{i'e}(n_e)$ if (i, i') $\in N \times N$, i.e. the cost function depends on the player himself and on the number of players that select the resource e.

Where: $n_e = |\{i : e = s_i\}|$, and $(n_1, n_2, ..., n_m)$ is called the congestion vector corresponding to a strategy $s = (s_1, s_2, ..., s_n)$.

This class of congestion games will be used in our approach to model and resolve the clustering problem.

*Nash Equilibrium in SCGPSC*

Malchtaich proved that this sub-class of congestion games does not generally admit a potential [15], so, it does not always converge to Nash equilibrium as it may be cyclic.
However, Matchtaich proved that SCGPSC possess always a Nash equilibrium. He used the proof by induction on the number *n* of players, where it is supposed that an instance of the game with n-1 players has a Nash equilibrium then the game with n players has also a Nash equilibrium. For the complete proof see [14]. This proof is constructive and implicitly describes an efficient algorithm for finding an equilibrium for a given n-player SCGPSC game, by adding one player after the other in at most $\binom{n+1}{2}$ steps [14].

### III. PROPOSED APPROACH

Our main contribution is the design of a multi-objective clustering algorithm based on game theory with parallel (or simultaneous move) games. We use the class of non-cooperative simultaneous-move game presented above; called Singleton Congestion game with Player-Specific Cost (SCGPSC), to model the clustering problem, where the players are part of the initial dataset, the resources are the rest of the dataset, and the cost function is an optimisation function of two conflicting objectives: connectedness and separation.

*A. Optimisation Objectives*

The main goal is to optimise a problem that consists of two conflicting optimisation objectives, the first is R-Square, and the other is the connectivity of the clusters based on the Euclidean distance. The combination of those two objectives was first used by Heloulou et al. [16] in a clustering algorithm that uses sequential game theory. R-Square [17], [18] is optimal when the number of clusters is high and the connectivity of clusters is optimal when the number of clusters is low, so, the compromise of those two conflicting objectives guarantees a good quality clustering according to the experiments presented in [16].

R-square is given by the formula:

$$R^2(C) = \frac{I_R(C)}{I_A(C) + I_R(C)} \quad (1)$$

Where $I_R(C)$ is inter-cluster inertia which measures the separation of the clusters [19]:

$$I_R(C) = \frac{1}{n} \sum_{i=1}^{K} |C_i| * d^2(ch_i, g), \quad (2)$$

Where $ch_i$ is the cluster-head of $C_i$, $g = (g_1, ..., g_m)$ and $g_j$ is the gravity centre of the dataset along the $j^{th}$ dimension.

$$g_j = \frac{1}{n} \sum_{i=1}^{n} o_{ij}. \quad (3)$$

$I_A(C)$ is the intra-cluster inertia which should be as weak as possible to have a set of homogeneous clusters. It is given as [19]:

$$I_A(C) = \frac{1}{n} \sum_{C_i \in C} \sum_{\omega \in C_i} d^2(\omega, ch_i). \quad (4)$$

Where: $chi$ is the cluster-head of the cluster $C_i$.

The connectivity measure is intended to assign similar data to the same cluster. It is given by the formula:

$$Connc(C) = \sum_{i=1}^{K} \frac{|C_i|}{n} * Connc(C_i), \quad (5)$$

Where $Connc(C_i)$ is the connectivity of the cluster $C_i$, and it is given by

$$Connc(C_i) = \frac{1}{|C_i|} \sum_{h=1}^{|C_i|} \frac{\sum_{j=1}^{L} \chi_{h, nn_{hj}}}{L}, \quad (6)$$

Where

$$\chi_{r,s} = \begin{cases} 1, & if \ r, s \ \epsilon \ C_i \\ 0, & otherwise \end{cases}$$

$nn_{hj}$ is the $j^{th}$ nearest neighbour of object h, and L is a

parameter indicating the number of neighbours that contribute to the connectivity measure. The connectivity value should be maximised. Heloulou et al. [16] have defined the product $\varphi(C)$ that combines the two objectives and should be maximised:

$$\varphi(C) = R^2(C) * Connc(C) \qquad (7)$$

### B. The Model

We model the clustering problem as a single-act nonzero-sum multi-player singleton congestion game with player-specific cost function. Hence, the game G is defined by the tuple:

$$G = (N, E, (X_k)_{k \in N}, (co_{ek})_{e \in E, k \in N}) \qquad (8)$$

Where
$N = \{ ch_i | ch_i = \underset{k: o_k \in O_i}{argmin} \, dm(o_k),$
$O_i = O_{i-1} \setminus \{ch_{i-1}, o_j \, | \, dis[j][ch_{i-1}] < \frac{Dmax}{n_0}\},$
$i = 0..n_0 - 1, j = 0..|O_{i-1}|\}$

Where: N is the set of players, i.e. objects with more density around them, $dm$ is the dissimilarity of an object, $Dmax$ is the maximum distance between two objects in the dataset, and $n_0$ is the initial number of players.

$E = \{o : o \in O \setminus N\}$
$X_k = \{\{o\} | o \in E\}, k = 0..|N|-1;$
$co_{ek}(n_e)$
$= \begin{cases} -\left(Connec(C_k^{(t)}) * R^2(C^{(t)})\right), k = 0..|N|-1, if\ n_e = 1 \\ \infty \quad, if\ n_e > 1 \end{cases}$
$, e \in E, n_e = |\{z: e = x_z\}|$

And the payoff of player k for strategy profile x is negative cost:

$$u_k(x) = -co_k(x)$$

### C. Model Implementation

In this section, we explain in detail our approach of clustering and we present our new algorithm of multi-objective clustering; called MOCA-SM.

MOCA-SM consists of three main steps; the first step consists of the identification of the players of the MOCA-SM game. The second step is the identification of the composition of the initial clusters. The third step is the refinement of the final clusters by merging the clusters resulted from the previous step. This is summarised in Algorithm 1.

---

**Algorithm 1:** MOCA-SM

**Inputs:** Dataset O=$\{o_0, o_1, ..., o_{m-1}\}$, number of final clusters f
**Outputs:** set of clusters

1: t←0
2: compute distance matrix DIS between all objects of O
3: **Algorithm 2:** Identification of initial players, the set ch
4: **repeat**
5:    t←t+1
6:    **Algorithm 3:** construct the game in formula (8)
7:    **Algorithm 4:** compute Nash equilibrium
8:    **for** each player
9:      **if** Nash equilibrium enhance the clustering then
10:        Allocate the Nash equilibrium strategy for player
11:      **else**
12:        Player out of game
13:      **end if**
14:    **end for**
15: **until** (O is empty, or all players are out of the game)
16: **Algorithm 5:** Merge the resulted clusters
17: assign the left objects to the closest cluster

---

**Algorithm 2:** Identification of initial players

**Inputs:** Dataset O=$\{o_0, o_1, ..., o_{m-1}\}$
**Outputs:** Set ch of initial cluster-heads (players) ch=$\{ch_1, ch_2, ..., ch_k\}$

1: compute $n_0$, the initial number of players using formula (9)
2: compute dissimilarity of each object $o_i$ of O: $dm(o_i)$ using formula (10)
3: O'← O
4: **repeat**
5:    Compute the object $o_x$ with the most density around in O'
6:    Add $o_x$ to ch
7:    Eliminate $o_x$ from the O'
8:    Eliminate close objects to $o_x$ from the O'
9: **until** (O' is empty or |ch| = $n_0$)

---

**Algorithm 3:** Formulation of the game

**Inputs:** players ch = $\{ch_1, ch_2, ..., ch_k\}$, dataset O=$\{o_0, o_1, ..., o_l\}$,
**Outputs:** Matrix *Cost* containing the costs of all strategies to all players

1: Identify available strategies over the set O
2: **for each** player $\in$ ch
3:    **for each** strategy
4:      **for each** value of the congestion vector
5:        compute the cost of the singleton strategy for the player using formula
6:        assign the cost to the matrix Cost
7:      **end for**
8:    **end for**
9: **end for**

---

**Algorithm 4:** Merging close clusters

**Inputs:** set of initial clusters set C=$\{c_0, c_1, ..., c_{n0-1}\}$, the number of the final clusters f
**Outputs:** set of final clusters C

1: compute the distance between all initial clusters
2: **repeat**
3:    merge the two closest clusters
4: **until the number of clusters is reached**

TABLE 1. DATASETS DESCRIPTION

| Datasets | | Instances | Attributes | Clusters |
|---|---|---|---|---|
| **Synthetic datasets** | **Spharical_3_4** | 400 | 3 | 4 {100, 100, 100, 100} |
| | **Dataset_3_2** | 76 | 2 | 3 {13, 43, 20} |
| | **Spiralsquare** | 1500 | 2 | 6 {116, 134, 125, 125, 500, 500} |
| **Real-world datasets** | **Iris** | 150 | 4 | 3 {50, 50, 50} |
| | **Wine** | 178 | 13 | 3 {59, 71, 48} |

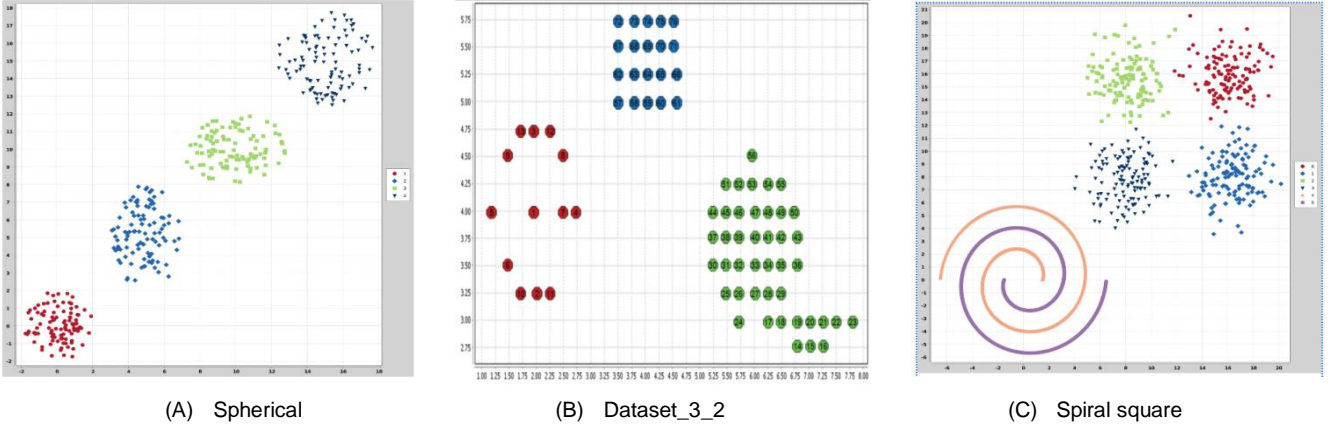

(A) Spherical  (B) Dataset_3_2  (C) Spiral square

Fig. 1. Synthetic datasets, (a)Spharical_3_4 (b) Dataset_3_2 (c) Spiral square

*Step1: Identification of the initial players*

This step aims at the identification of the players (cluster heads). The initial number of clusters is estimated using the formula:

$$n_0 = m/L \qquad (9)$$

Where $m$ is the size of the dataset, and $L$ is the number of neighbours used to find the connectivity of each cluster. The dissimilarity of each object is then compared to all other objects in the dataset, and it is computed as follows:

$$dm(o) = \frac{1}{m-1}\sum_{o_i \in O, o_i \neq o} dis[o][o_i] \qquad (10)$$

Objects with the smallest *dm* values are the objects with the most density around them; consequently, they are chosen to be cluster-heads (or players).

*Step2: Establishment of initial clusters' composition*

This step aims at determining the composition of the initial clusters by constructing the game described in formula (8). The players of the game are the objects with the highest density around them. This game is played several times until all objects are allocated or when no player wants to play again. Algorithm 3 summarises this step.

*Step 3: Merging close clusters*

After having $n_0$ initial clusters from step2, in this step, we perform a merging operation until we obtained very distinct clusters. First, the distance between each pair of the initial clusters is calculated, then, the most two closest clusters are merged within a reasonable distance based on compactness.

## IV. EXPERIMENTATIONS AND CLUSTERING VALIDATION

### A. Settings

Experiments were conducted on 2.50GHz Intel ® Core ™ i5-3210M with 8 GB of RAM. All the algorithms were implemented in Java programming language. We compared our approach with well-known clustering algorithms. These include K-means, DBSCAN (density-based spatial clustering of applications with noise), and SOM (Self-organizing Map). The source code for those algorithms was taken from Java Machine Learning Library JavaML [1].

The parameter $L$ is the number of neighbours required to define the connectivity. This parameter depends on the size of the data and its dispersion. In the datasets used in our experiments, $L$ is set to value 9 when the size of the dataset is less than 150; the value 14 when the size of the dataset is between 150 and 500; the value 28 when the dataset is bigger than 500. The experiments were conducted on five datasets; three of them are synthetic: Spharical_3_4[2], Dataset_3_2, and Spiral square[3] (see Fig. 1), the other two are real-world: Iris[4] and Wine[5], which are well-known datasets for clustering evaluation (see TABLE 1).

---

[1] http://java-ml.sourceforge.net/
[2] Available at GitHub : https://github.com/deric/clustering-benchmark/tree/b47cdbb7028a61d632e2c63901f868e99444b350
[3] Available at GitHub : https://github.com/deric/clustering-benchmark/tree/b47cdbb7028a61d632e2c63901f868e99444b350
[4] Available at UCI Machine Learning Repository. : ftp://ftp.ics.uci.edu/pub/machine-learning-databases
[5] Available at UCI Machine Learning Repository. : ftp://ftp.ics.uci.edu/pub/machine-learning-databases

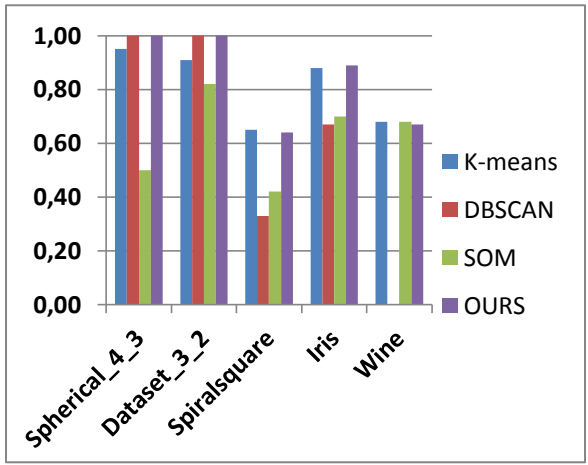

Fig. 2. Purity values resulted from various Algorithms on different datasets. A high value of purity indicates a better clustering result.

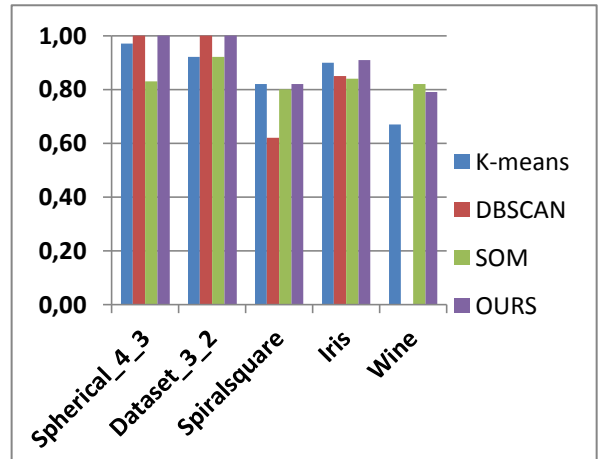

Fig. 3. Rand Index resulted from running various Algorithms on different datasets. A high value of Rand Index indicates a better clustering result.

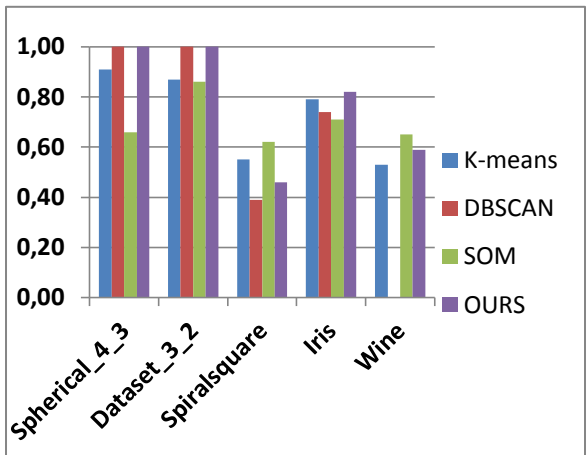

Fig. 4. F-measure resulted from running various Algorithms on different datasets. A high value of F-measure indicates a better clustering result.

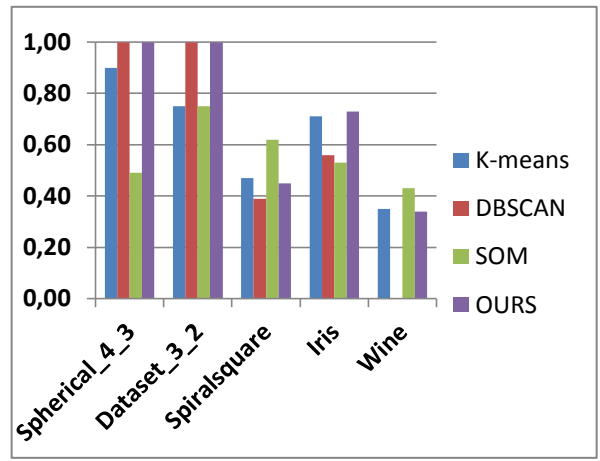

Fig. 5. ARI resulted from running various Algorithms on different datasets. A high value of ARI indicates a better clustering result.

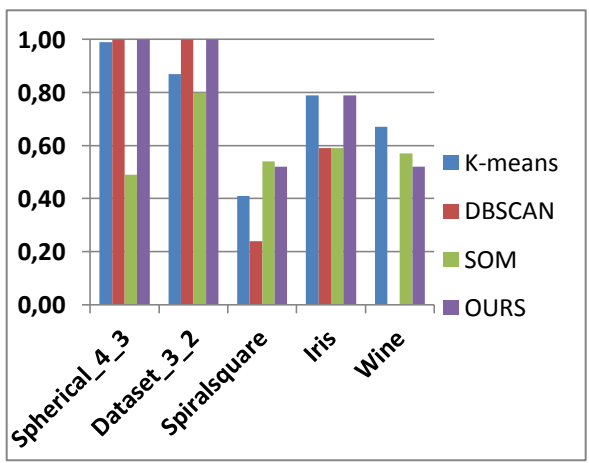

Fig. 6. Precision resulted from running various Algorithms on different datasets. A high value of Precision indicates a better clustering result.

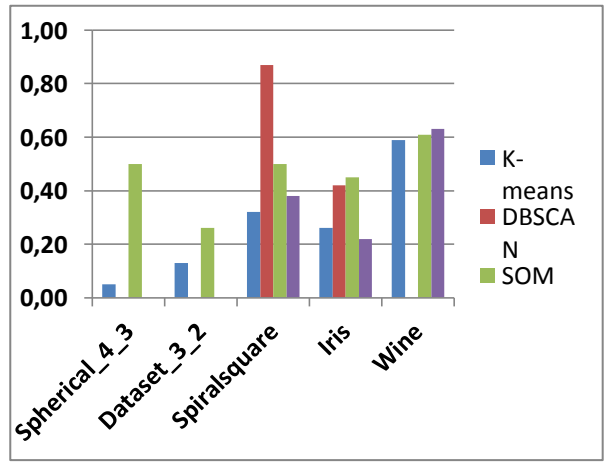

Fig. 7. Entropy resulted from running various Algorithms on different datasets. A low value of Entropy indicates a better clustering result.

In the aim of validating our approach, we use the following metrics:

- **Purity:** purity is the percentage of the objects that were classified correctly:
$$P_j = \frac{1}{n_j} \text{MAX}_i(n_j^i) \quad (11)$$

Where $n_j$ is the size of the cluster j, and $n_j^i$ is the number of correctly assigned objects. The overall purity of a clustering is given by:
$$P = \sum_{j=1}^{k} \frac{n_j}{n} P_j \quad (12)$$

- **Rand Index (RI):** The Rand index measures the percentage of correct decisions. It is calculated as follows [20]:
$$RI = \frac{TP+TN}{TP+FP+FN+TN} \quad (13)$$

Where TP is a true positive, TN is a true negative, FP is a false positive, and FN is a false negative.

- **Adjusted Rand Index:**[21] is the corrected-for-chance version of the Rand index, it is given by the following:
$$ARI = \frac{\sum_{lk} \binom{n_{lk}}{2} - [\sum_l \binom{n_l}{2} * \sum_k \binom{n_k}{2}] / \binom{n}{2}}{\frac{1}{2}[\sum_l \binom{n_l}{2} * \sum_k \binom{n_k}{2}] - [\sum_l \binom{n_l}{2} * \sum_k \binom{n_k}{2}] / \binom{n}{2}} \quad (14)$$

- **F-measure:** F-Measure provides a single score that balances both the concerns of precision and recall in one number, it is given by the formula [22]:
$$F_w = \frac{(W^2+1)*P*R}{W^2*P+R}, \quad (15)$$
$$E_j = -\frac{1}{\log k} \sum_{i=1}^{k} \frac{n_j^i}{n_j} \text{LOG} \frac{n_j^i}{n_j} \quad (16)$$

Where k is the number of clusters in the dataset, w is a positive real value, P is the precision, and R is the recall.
$$P = \frac{TP}{TP+FP} \quad (17)$$
$$R = \frac{TP}{TP+FN} \quad (18)$$

- **Entropy:** the entropy shows how the various classes of objects are distributed within each cluster [23]. It is given by the following formula:
$$Entropy = \sum_{j=1}^{K} \frac{n_j}{n} E_j \quad (19)$$

Where $E_j$ is the entropy of cluster j, it is given as follows:
$$E_j = -\frac{1}{\log k} \sum_{i=1}^{k} \frac{n_j^i}{n_j} \log \frac{n_j^i}{n_j} \quad (20)$$

and k is the number of clusters in the dataset.

*B. Results and Discussion*

The experimental results indicate that our approach, MOCA-SM, obtains good results for all datasets overall (see Fig. 2, Fig. 3, Fig. 4, Fig. 5, Fig. 6, and Fig. 7). One can notice that other approaches returned good results for some datasets and average results in others. In the following we discuss the results of each dataset.

(1) **Spharical_3_4:** for this dataset, our approach returned perfect clustering, as it is shown in the Figures above.
While K-means gets noticeably average results, SOM gets the lowest results for this dataset in all evaluation metrics used. DBSCAN did very well too.

(2) **Dataset_3_2**: Similar to the previous dataset, our approach and DBSCAN performed a perfect clustering, while K-means and SOM produced average quality clustering. This is for all the evaluation metrics used.

(3) **Spiral square:** for this challenging synthetic dataset, all the four algorithms performed worse than on the first two datasets. Although k-means is slightly better than our approach, they both obtained better results in terms of cluster purity, Rand Index, and entropy than DBSCAN and SOM, which returned less than average results for this dataset. On the other hand, SOM obtained the higher precision and F-measure results, closely followed by our approach.

(4) **Iris:** for this real-world dataset, our approach returned the best results in all evaluation metrics, closely followed by k-means, then DBSCAN and SOM. Overall, these four approaches high quality results for this dataset.

(5) **Wine:** for this dataset, SOM got higher results, closely followed by our approach for all metrics except for the precision where k-means takes the lead. The DBSCAN results were not taken into account because it eliminated 80% of the dataset points considering them as noise.

The results of this comparative study for each dataset and metric are presented in Fig. 2, Fig. 3, Fig. 4, Fig. 5, Fig. 6, and Fig. 7.

## V. CONCLUSION AND FUTURE WORK

In this paper, we presented congestion games with player-specific functions that model a clustering problem into a game. The use of game theory tools provides a solid mathematical background of the proposed solution. This allows us to understand how the clustering is done and the results can be easily interpreted. In addition, the propose approach returned very good clustering results, as shown in the previous section on various datasets and metrics.

The proposed approach consists of three main phases: in the first phase we identify the players over a set of objects or data points, which are considered as the centres of the clusters. In the second phase, each player aims to improve their gains in terms of connectivity and R-square. Each player (or cluster-head) plays and attempts to reach Nash equilibrium. In the third phase, when all players stop playing, the merging starts. Two clusters are merged if they are very close to each other.

Much further work is needed, especially in the 3$^{rd}$ phase to give our approach the ability of automatically deciding on the number of the final clusters.

Although the scalability of the approach has not been presented in this paper, some preliminary tests have been conducted and the results very promising in this matter. As continuation of this work, we will study its complexity and effectiveness in details on heterogeneous datasets in terms of quality, noise and outliers. Moreover, we will extend the approach to categorical attributes and compare it to the best algorithms for that category of attributes [24].

ACKNOWLEDGMENT

The authors wish to thank the Insight Centre for Data Analytics, UCD, Dublin, for inviting the first author as a visiting researcher while she was working on the development of this clustering approach.